\documentclass[aps,pra,showpacs,showkeys,preprint]{revtex4}
\usepackage{amstext}
\usepackage{amsmath}
\usepackage{graphicx}
\usepackage{bm}

\def\qed{\leavevmode\unskip\penalty9999 \hbox{}\nobreak\hfill
     \quad\hbox{\leavevmode  \hbox to.77778em{%
               \hfil\vrule   \vbox to.675em%
               {\hrule width.6em\vfil\hrule}\vrule\hfil}}
     \par\vskip3pt}

\newtheorem{theorem}{Theorem}

\newtheorem{lemma}[theorem]{Lemma}
\newtheorem{cor}[theorem]{Corollary}

\begin{document}
\title{Entanglement conditions for multi-mode states}
\vspace{2ex}

\author{
Zong-Guo Li$^1$, Shao-Ming Fei$^{1, 2}$, Zhi-Xi Wang$^1$ and Ke Wu$^1$}
\affiliation{
$^1$ Department of Mathematics, Capital Normal University, Beijing 100037, China\\
$^2$ Max Planck Institute for Mathematics in the Sciences, D-04103 Leipzig, Germany}

\begin{abstract}
We provide a class of inequalities for detecting entanglements in
multi-mode systems. Necessary conditions for fully separable,
bi-separable and sufficient conditions for fully entangled states are
explicitly presented.
\end{abstract}
\maketitle
\section{Introduction}
Entanglement plays a key role in the rapidly developing field of
quantum information processing. Consequently, the study of
entanglement of bi- and multipartite systems has been the focus of
research in quantum information theory. In recent years,
increasing attention was paid to infinite dimensional systems, the
so called continuous-variable (CV) systems. In particular, Gaussian
state entanglement, a special case of CV systems, has aroused
great interest\cite{prl96,simon,lmd,ferraro,giedke,pra64,hyllus,serafini}.
The quantum teleportation network has been recently demonstrated
experimentally with the use of fully symmetric three-mode Gaussian
states\cite{yonezawa}. Many efforts\cite{peter,giedke,hofmann,
guhne,simon,lmd,Hillery,shchukin,shchukin3,horodecki,
ferraro,serafini,braunstein,hyllus} have been devoted to the study
of detecting entanglement for CV systems.

It is of particular relevance to provide theoretical methods to
determine the entanglement of a CV quantum system in quantum
information processing and computation. Such an interest stems
from the practical needs in the implementations of realistic
information protocols. There has been some progress in the
criteria for entanglement by uncertainty
relations\cite{Hillery,shchukin,shchukin3,hofmann,guhne,peter}.
In \cite{Hillery} Hillery and Zubairy have provided a class of
inequalities for detecting entanglement of two-mode states.

In this paper, we generalize the inequalities to the case of
multi-mode states. We derive a class of inequalities for detecting
the entanglement of multi-mode states and some inequalities to
detect fully entangled states. These quantities and inequalities
can be, in principle, measured in standard homodyne correlation
experiments\cite{shchukin2}.

\section{Bipartite entangled conditions for 3-mode states}
Firstly we consider the case of three-mode states. We focus on
three-mode harmonic oscillators. The derived conditions can also
be applied to radiation modes, motional states of trapped ions,
and related systems.

Let operators $a$, $b$ and $c$ be the
annihilation operators of the first ($A$), the second ($B$) and
the third ($C$) mode, respectively. Set $L_{1}=abc^{\dagger
}+a^{\dagger }b^{\dagger }c$ and $L_{2}=i(abc^{\dagger
}-a^{\dagger }b^{\dagger }c)$. We have
\begin{eqnarray}
\left[ L_{1},L_{2}\right]
&=&2i\{N_{a}N_{b}-N_{a}N_{c}-N_{b}N_{c}-N_{c}\},
\end{eqnarray}%
and
\begin{eqnarray}
\label{ineq02} (\Delta L_{1})^{2}+(\Delta L_{2})^{2} &=&4\langle
N_{a}N_{b}N_{c}\rangle -4\left\vert \langle abc^{\dagger
}\rangle \right\vert ^{2}\nonumber \\
&&+2(\langle N_{a}N_{b}\rangle +\langle N_{b}N_{c}\rangle +\langle
N_{a}N_{c}\rangle +\langle N_{c}\rangle ),
\end{eqnarray}
where $N_{a}=a^{\dagger }a,N_{b}=b^{\dagger }b$ and
$N_{c}=c^{\dagger }c$. For a pure state that is $AB|C$ bi-separable
with respect to the first, second and third mode, we have
\begin{eqnarray}
(\Delta L_{1})^{2}+(\Delta L_{2})^{2}&=&4\langle N_{a}N_{b}\rangle
\langle N_{c}\rangle -4\left\vert \langle ab\rangle \langle
c^{\dagger }\rangle \right\vert ^{2}\nonumber \\
&&+2(\langle N_{a}N_{b}\rangle +\langle N_{b}N_{c}\rangle +\langle
N_{a}N_{c}\rangle +\langle N_{c}\rangle ).
\end{eqnarray}%
Noting that the Schwarz inequality implies that, for any state $|\Psi
\rangle$, $\left\vert \langle a\rangle \right\vert ^{2}=\left\vert
\langle \Psi \left\vert a\right\vert \Psi \rangle \right\vert
^{2}\leq |\langle \Psi ||^{2} \,\cdot |a\left\vert \Psi \rangle \right\vert
^{2}=\langle \Psi |a^{\dagger }a|\Psi \rangle =\langle N_{a}\rangle$,
$|\langle ab\rangle |^{2}\leq \left\langle N_{a}N_{b}\right\rangle$,
for such a pure $AB|C$ separable state we get
\begin{equation}
\label{part_c} (\Delta L_{1})^{2}+(\Delta L_{2})^{2}\geq 2(\langle
N_{a}N_{b}\rangle +\langle N_{b}N_{c}\rangle +\langle
N_{a}N_{c}\rangle +\langle N_{c}\rangle ) .
\end{equation}

\begin{lemma}For any density matrix
\begin{equation}
 \label{ineq05} \rho =\sum_{k}p_{k}\rho
_{k},~~~0<p_k\leq 1,~~~\sum_i p_i =1,
\end{equation}and a variable $S$,we have
\begin{equation}
\label{genin} (\Delta S)^{2}\geq \sum_{k}p_{k}(\Delta S_{k})^{2}.
\end{equation}
\end{lemma}

The lemma can be proved by the convexity. From the lemma, we have
that the inequality (\ref{part_c}) holds for $AB|C$ separable
\underline{mixed} states as well.

On the other hand the uncertainties $\Delta L_{1}$ and $\Delta
L_{2}$ also satisfy
\begin{eqnarray}
\label{genin2} (\Delta L_{1})^{2}+(\Delta L_{2})^{2} &\geq &2\Delta
L_{1}\Delta L_{2}\geq
|\left\langle \left[ L_{1},L_{2}\right] \right\rangle |  \nonumber \\
&=&2|\langle N_{a}N_{c}\rangle -\langle N_{b}N_{c}\rangle -\langle
N_{a}N_{c}\rangle -\langle N_{c}\rangle |.
\end{eqnarray}
Comparing this result, which holds for any state, with inequality
(\ref{part_c}), which holds for any $AB|C$ separable states, we
see that the right-hand side of inequality (\ref{genin2}) is
always less than or equal to that of inequality (\ref{part_c}).
Consequently, there may be states that violate inequality
(\ref{part_c}) while satisfying inequality (\ref{genin2}). Thus we
have the following theorem:

\begin{theorem}
If the following inequality holds for a state $\rho$,
\begin{equation*}
(\Delta L_{1})^{2}+(\Delta L_{2})^{2}< 2(\langle N_{a}N_{b}\rangle
+\langle N_{b}N_{c}\rangle +\langle N_{a}N_{c}\rangle +\langle
N_{c}\rangle ),
\end{equation*}then the state is $AB|C$ entangled.
\end{theorem}
For a state $|\Psi \rangle =(|0\rangle _{a}|0\rangle _{b}|1\rangle
_{c}+|0\rangle _{a}|1\rangle _{b}|0\rangle _{c}+|1\rangle
_{a}|0\rangle _{b}|1\rangle _{c}+|1\rangle _{a}|1\rangle
_{b}|0\rangle _{c})/2,$ we have
\begin{eqnarray}
(\Delta L_{1})^{2}+(\Delta L_{2})^{2}< 2(\langle N_{a}N_{c}\rangle
+\langle N_{a}N_{b}\rangle +\langle N_{b}N_{c}\rangle +\langle
N_{c}\rangle ).\nonumber
\end{eqnarray}
Therefore the theorem shows that the state $|\Psi \rangle $ is
$AB|C$ entangled.

Comparing (\ref{ineq02}) with (\ref{part_c}) we have the following
corollary:
\begin{cor}If the following inequality holds for a state $\rho$,
\begin{equation}
\label{part_c2} \langle N_{a}N_{b}N_{c}\rangle <\left\vert \langle
abc^{\dagger }\rangle \right\vert ^{2},
\end{equation}then the state is $AB|C$ entangled.
\end{cor}

Note that the Schwarz inequality implies that
\begin{equation}
\label{ineq09} \left\vert \langle abc^{\dagger }\rangle \right\vert
^{2}\leq \langle N_{a}N_{b}(N_{c}+1)\rangle.
\end{equation}
Formulae (\ref{part_c2}) and (\ref{ineq09}) suggest that there is
a family of similar conditions for detecting entanglement, if one
considers operator $a^{m}b^{n}(c^{\dag})^{l}$ instead of
$abc^{\dag}$. So we have the following theorem:
\begin{theorem}
If the following inequality holds for a state $\rho$,
\begin{equation}
\label{ineq14} |\langle a^{m}b^{n}(c^{\dag})^{l}\rangle |^{2}>
\langle
(a^{\dag})^{m}a^{m}(b^{\dag})^{n}b^{n}(c^{\dag})^{l}c^{l}\rangle,
\end{equation}
where $m$, $n$ and $l$ are positive integers, then $\rho$ is $AB|C$ entangled.
\end{theorem}

{\em Proof:} For a $AB|C$ separable pure state we have
\begin{eqnarray}
\label{ineq10} |\langle a^{m}b^{n}(c^{\dag })^{l}\rangle
|^{2}=|\langle a^{m}b^{n}\rangle |^{2}|\langle c^{l}\rangle
|^{2}&\leq &\langle (a^{\dag })^{m}a^{m}(b^{\dag })^{n}b^{n}\rangle
\langle (c^{\dag })^{l}c^{l}\rangle \nonumber \\
&=&\langle (a^{\dag
})^{m}a^{m}(b^{\dag })^{n}b^{n}(c^{\dag })^{l}c^{l}\rangle.
\end{eqnarray}
To show that formula (\ref{ineq10}) is valid for $AB|C$ separable
mixed state (\ref{ineq05}), defining $\textsc{a}=a^{m}$,
$\textsc{b}=b^{n}$ and $\textsc{c}=c^{l}$, we have
\begin{eqnarray}
|\langle \textsc{a}\textsc{b}\textsc{c}^{\dag }\rangle | &=&|\sum_{k}p_{k}
tr(\rho _{k}\textsc{a}\textsc{b}\textsc{c}^{\dag })|
\leq \sum_{k}p_{k}|tr(\rho _{k}\textsc{a}\textsc{b}\textsc{c}^{\dag })|  \nonumber \\
&=&\sum_{k}p_{k}|tr(\rho _{k}^{(ab)}\textsc{a}\textsc{b})_{k}tr(\rho _{k}^{(c)}\textsc{c}^{\dag
})| \leq \sum_{k}p_{k}\langle \textsc{a}^{\dag }\textsc{a}\textsc{b}^{\dag }\textsc{b}\rangle
^{1/2}_{k}\langle \textsc{c}^{\dag }\textsc{c}\rangle ^{1/2}_{k}.
\end{eqnarray}%
From the convexity, we have that
\begin{eqnarray}
|\langle \textsc{a}\textsc{b}\textsc{c}^{\dag }\rangle | &\leq &
(\sum_{k}p_{k}\langle \textsc{a}^{\dag }\textsc{a}\textsc{b}^{\dag }\textsc{b}\rangle_{k} \langle
\textsc{c}^{\dag}\textsc{c}\rangle_{k} )^{1/2}  \nonumber \\
&=&(\sum_{k}p_{k}\langle \textsc{a}^{\dag }\textsc{a}\textsc{b}^{\dag }\textsc{b}\textsc{c}^{\dag }\textsc{c}\rangle_{k}
)^{1/2} =\langle \textsc{a}^{\dag }\textsc{a}\textsc{b}^{\dag }\textsc{b}\textsc{c}^{\dag }\textsc{c}\rangle ^{1/2},
\end{eqnarray}
which shows that the inequality (\ref{ineq10}) does indeed hold
for all $AB|C$ separable states. This ends the proof.\qed

In analogy to $L_{1}$ and $L_{2}$, we have the following theorems:
\begin{theorem}
If any one of the following inequalities hold for a state $\rho$,
\begin{equation}
\label{ineq15}
 (\Delta J_{1})^{2}+(\Delta J_{2})^{2}< 2(\langle
N_{a}N_{c}\rangle +\langle N_{a}N_{b}\rangle +\langle
N_{b}N_{c}\rangle +\langle N_{a}\rangle ),
\end{equation}
\begin{equation}
\label{ineq17} |\langle (a^{\dag })^{m}b^{n}c^{l}\rangle
|^{2}>\langle (a^{\dag })^{m}a^{m}(b^{\dag })^{n}b^{n}(c^{\dag
})^{l}c^{l}\rangle,
\end{equation}where $J_{1}=ab^{\dag
}c^{\dag }+a^{\dag }bc$, $J_{2}=i(-ab^{\dag }c^{\dag }+a^{\dag
}bc)$, $m$, $n$ and $l$ are positive integers, then the state is entangled between
A-party and BC-party.
\end{theorem}
This theorem can be directly proved by using the techniques in proving
theorems 2 and 4.

We can Similarly obtain the following theorem for entangled states
between B-party and AC-party:
\begin{theorem}
If any one of the following inequalities hold for a state $\rho$,
\begin{equation}
\label{ineq18} (\Delta K_{1})^{2}+(\Delta K_{2})^{2}< 2(\langle
N_{a}N_{c}\rangle +\langle N_{a}N_{b}\rangle +\langle
N_{b}N_{c}\rangle +\langle N_{b}\rangle).
\end{equation}
\begin{equation}
\label{ineq19} |\langle a^{m}(b^{\dag })^{n}c^{l}\rangle
|^{2}>\langle (a^{\dag })^{m}a^{m}(b^{\dag })^{n}b^{n}(c^{\dag
})^{l}c^{l}\rangle ,
\end{equation}where $K_{1}=ab^{\dag }c+a^{\dag }bc^{\dag }$ and $%
K_{2}=i(ab^{\dag }c-a^{\dag }bc^{\dag })$, $m$, $n$ and $l$ are positive integers,
then the state is $B|AC$ entangled .
\end{theorem}

In fact, there is another class of inequalities detecting the
bipartite entangled states. We have the following theorem:
\begin{theorem}
For a state $\rho$,

if $|\langle a^{m}b^{n}c^{l}\rangle |> \left[ \langle (a^{\dag
})^{m}a^{m}\rangle \langle (b^{\dag })^{n}b^{n}(c^{\dag
})^{l}c^{l}\rangle \right] ^{1/2}, $ then the state is $A|BC$
entangled;

if $|\langle a^{m}b^{n}c^{l}\rangle |> \left[ \langle(b^{\dag
})^{n}b^{n} \rangle \langle (a^{\dag })^{m}a^{m}(c^{\dag
})^{l}c^{l}\rangle \right] ^{1/2},$ then the state is $B|AC$
entangled;

if
$ |\langle a^{m}b^{n}c^{l}\rangle |> \left[ \langle (a^{\dag
})^{m}a^{m}(b^{\dag })^{n}b^{n}\rangle \langle (c^{\dag
})^{l}c^{l}\rangle \right] ^{1/2}, $ then the state is $AB|C$
entangled .
\end{theorem}

 {\em Proof:} We prove the condition for $A|BC$
entanglement.

From Schwarz inequality, we have
\begin{equation}
\label{ineq43} |\langle a^{m}b^{n}c^{l}\rangle |\leq \left[
\langle (a^{\dag })^{m}a^{m}\rangle \langle (b^{\dag
})^{n}b^{n}(c^{\dag })^{l}c^{l}\rangle \right] ^{1/2}
\end{equation} for a pure $A|BC$
separable state.

For $A|BC$ separable mixed state $\rho =\sum_{k}p_{k}\rho _{k}$,
where $\rho _{k}$ is the density matrix corresponding to a pure
$A|BC$ separable state, and $p_{k}$\ is the probability of $\rho
_{k},$ we have
\begin{eqnarray}
\label{ineq21}
|\langle \textsc{a}\textsc{b}\textsc{c}\rangle |^{2}
&=&|\sum_{k}p_{k}tr(\rho _{k}\textsc{a}\textsc{b}\textsc{c})|^{2}\leq
(\sum_{k}p_{k}|tr(\rho _{k}\textsc{a}\textsc{b}\textsc{c})|)^{2}  \nonumber \\
&=&\sum_{i,j}p_{i}p_{j}|\langle \textsc{a}\rangle _{i}\langle \textsc{b}\textsc{c}\rangle
_{i}||\langle
\textsc{c}^{\dag }\textsc{b}^{\dag }\rangle _{j}\langle \textsc{a}^{\dag }\rangle _{j}|  \nonumber \\
&\leq &\sum_{i,j}p_{i}p_{j}(\langle \textsc{a}^{\dag }\textsc{a}\rangle _{i}\langle
\textsc{b}^{\dag }\textsc{b}\textsc{c}^{\dag }\textsc{c}\rangle _{i}\langle
\textsc{b}^{\dag }\textsc{b}\textsc{c}^{\dag }\textsc{c}\rangle_{j}\langle \textsc{a}^{\dag }\textsc{a}\rangle _{j})^{1/2},
\end{eqnarray}
where $\textsc{a}=a^{m}$, $\textsc{b}=b^{n}$ and $\textsc{c}=c^{l}$. Set $ \langle \textsc{a}^{\dag
}\textsc{a}\rangle_{i}=x_{i}$, $\langle \textsc{b}^{\dag }\textsc{b}\textsc{c}^{\dag }\textsc{c}\rangle_{i}=y_{i}$.
(\ref{ineq21}) can be expressed as
\begin{equation}
\label{ineq22} |\langle \textsc{a}\textsc{b}\textsc{c}\rangle |^{2}\leq
\sum_{k}p_{k}^{2}x_{k}y_{k}+2\sum_{i<j}p_{i}p_{j}(x_{i}y_{i}x_{j}y_{j})^{1/2}.
\end{equation}
While $\langle \textsc{a}^{\dag }\textsc{a}\rangle \langle \textsc{b}^{\dag }\textsc{b}\textsc{c}^{\dag }\textsc{c}\rangle
$ can be expressed as
\begin{equation}
\label{ineq23}
\langle \textsc{a}^{\dag }\textsc{a}\rangle \langle \textsc{b}^{\dag }\textsc{b}\textsc{c}^{\dag
}\textsc{c}\rangle
=\sum_{k}p_{k}^{2}x_{k}y_{k}+\sum_{i<j}p_{i}p_{j}(x_{i}y_{j}+x_{j}y_{i}).
\end{equation}
Noting that $x_{i}y_{j}+x_{j}y_{i}\geq
2(x_{i}y_{i}x_{j}y_{j})^{1/2},$ we see that from (\ref{ineq22})
and (\ref{ineq23}) the inequality (\ref{ineq43}) holds for all
$A|BC$ bipartite separable states. Hence if a state violates this
inequality (\ref{ineq43}), it must be bipartite entangled between
A-part and BC-part. The other conditions can be proved similarly.\qed

For $m=n=l=1$, formula (\ref{ineq43}) implies that for a $A|BC$
separable state
\begin{equation}
\label{ineq49} |\langle abc\rangle |^{2}\leq \langle N_{a}\rangle
\langle N_{b}N_{c}\rangle.
\end{equation}
For an arbitrary state, we have $|\langle abc\rangle |^{2}\leq
\left[ \langle N_{a}+1\rangle \langle N_{b}N_{c}\rangle \right]$.
Therefore there may be states that do violate the inequality
(\ref{ineq49}) and are $A|BC$ entangled at the same time.

\section{fully entangled conditions for 3-mode states}

We first introduce the notions of fully separable 3-mode states
and fully entangled ones in\cite{prl83,uffink}. A 3-mode state is
fully separable if and only if the state can be described as
mixtures of $\rho_{1}\otimes\rho_{2}\otimes\rho_{3}$. A 3-mode
entangled state is fully entangled if it is not bi-partite separable.
We have our main theorem:

\begin{theorem}
If the following inequality holds for a state $\rho$,
\begin{equation}
\label{ineqeq31} (\Delta K(\phi ))^{2}<1,
\end{equation}
where $K(\phi )=e^{i\phi}a^{\dag }b^{\dag }c^{\dag }+e^{-i\phi }abc$,
$\phi\in [0,\,2\pi]$, then the state is fully entangled.
\end{theorem}

 {\em Proof:}
\begin{eqnarray}
(\Delta K(\phi ))^{2} &=&e^{2i\phi }\langle (a^{\dag }b^{\dag }c^{\dag
}-\langle a^{\dag }b^{\dag }c^{\dag }\rangle )^{2}\rangle +e^{-2i\phi
}\langle (abc-\langle abc\rangle )^{2}\rangle  \nonumber \\
&&+\langle (a^{\dag }b^{\dag }c^{\dag }-\langle a^{\dag }b^{\dag }c^{\dag
}\rangle )(abc-\langle abc\rangle )\rangle\nonumber \\
&&+\langle (abc-\langle abc\rangle )(a^{\dag }b^{\dag }c^{\dag
}-\langle a^{\dag }b^{\dag }c^{\dag }\rangle )\rangle.
\end{eqnarray}%
As $e^{2i\phi }\langle (a^{\dag }b^{\dag }c^{\dag }-\langle a^{\dag
}b^{\dag }c^{\dag }\rangle )^{2}\rangle +e^{-2i\phi }\langle
(abc-\langle abc\rangle )^{2}\rangle $ is real, by using the Schwarz
inequality
\begin{eqnarray}
&&|\langle (abc-\langle abc\rangle )^{2}\rangle |\nonumber \\
&\leq& \lbrack \langle (abc-\langle abc\rangle )(a^{\dag }b^{\dag
}c^{\dag }-\langle a^{\dag }b^{\dag }c^{\dag }\rangle )\rangle
\langle (a^{\dag }b^{\dag }c^{\dag }-\langle a^{\dag }b^{\dag
}c^{\dag }\rangle )(abc-\langle abc\rangle )\rangle ]^{1/2},
\end{eqnarray}
we obtain
\begin{eqnarray}
\label{fullsep} &&(\Delta K(\phi ))^{2}\nonumber\\ &\geq &-|\langle
(a^{\dag }b^{\dag }c^{\dag }-\langle a^{\dag }b^{\dag }c^{\dag
}\rangle )^{2}\rangle |-|\langle (abc-\langle
abc\rangle )^{2}\rangle |  \nonumber \\
&&+\langle (a^{\dag }b^{\dag }c^{\dag }-\langle a^{\dag }b^{\dag }c^{\dag
}\rangle )(abc-\langle abc\rangle )\rangle +\langle (abc-\langle abc\rangle
)(a^{\dag }b^{\dag }c^{\dag }-\langle a^{\dag }b^{\dag }c^{\dag }\rangle
)\rangle  \nonumber \\
&\geq &-2[\langle (abc-\langle abc\rangle )(a^{\dag }b^{\dag }c^{\dag
}-\langle a^{\dag }b^{\dag }c^{\dag }\rangle )\rangle \langle (a^{\dag
}b^{\dag }c^{\dag }-\langle a^{\dag }b^{\dag }c^{\dag }\rangle )(abc-\langle
abc\rangle )\rangle ]^{1/2}  \nonumber \\
&&+\langle (a^{\dag }b^{\dag }c^{\dag }-\langle a^{\dag }b^{\dag }c^{\dag
}\rangle )(abc-\langle abc\rangle )\rangle +\langle (abc-\langle abc\rangle
)(a^{\dag }b^{\dag }c^{\dag }-\langle a^{\dag }b^{\dag }c^{\dag }\rangle
)\rangle  \nonumber \\
&=&[(\langle (N_{a}+1)(N_{b}+1)(N_{c}+1)\rangle -|\langle abc\rangle
|^{2})^{1/2}-(\langle N_{a}N_{b}N_{c}\rangle -|\langle abc\rangle
|^{2})^{1/2}]^{2}.
\end{eqnarray}%
The inequality (\ref{fullsep}) is valid for all states. In
particular, if the state is a fully separable pure state,
(\ref{fullsep}) becomes
\begin{eqnarray}
(\Delta K(\phi ))^{2}\geq [ (\langle N_{a}+1\rangle \langle
N_{b}+1\rangle \langle N_{c}+1\rangle -|\langle abc\rangle
|^{2})^{1/2} \nonumber \\
-(\langle N_{a}\rangle \langle N_{b}\rangle \langle N_{c}\rangle
-|\langle abc\rangle |^{2})^{1/2}]^{2}.
\end{eqnarray}

Denote $x_{1}=\langle N_{a}\rangle ,x_{2}=\langle N_{b}\rangle
,x_{3}=\langle N_{c}\rangle $ and $z=|\langle abc\rangle |^{2}.$ Let
us find the minimum of the function
\begin{equation}
F(x_{1},x_{2},x_{3})=\sqrt{(x_{1}+1)(x_{2}+1)(x_{3}+1)-z}-\sqrt{%
x_{1}x_{2}x_{3}-z},
\end{equation}%
in the region $x_{1}x_{2}x_{3}\geq z\geq 0.$

There are two cases: 1. $F(x_{1},x_{2},x_{3})$ has minimum in the
region of interest. From
$\partial F(x_{1},x_{2},x_{3})/\partial x_{i}=0$, $i=1,2,3$,
$F$ acquires the minimum at $x_{1}=x_{2}=x_{3}=x$,
\begin{equation}
F_{\min}=\sqrt{(x+1)^{3}-z}-\sqrt{x^{3}-z}.
\end{equation}%
As $(\sqrt{(x+1)^{3}-z})^{2}-(1+\sqrt{x^{3}-z})^{2} \geq
(x+1)^{3}-(x^{3}+1+2\sqrt{x^{3}}),$ we have $F_{\min}\geq1$ in the
region of interest.

2. $F(x_{1},x_{2},x_{3})$ has no local minimum in the region of
interest. Then the minimum of the function must lie on the
boundary, $x_{1}x_{2}x_{3}=z$ or at least one of the $x_{1}$, $x_{2}$ and $x_{3}$ goes
to infinity.

On the boundary $x_{1}x_{2}x_{3}=z$ we have
\begin{equation}
F(x_{1},x_{2},x_{3})=\sqrt{(x_{1}+1)(x_{2}+1)(x_{3}+1)-x_{1}x_{2}x_{3}}\geq
1.
\end{equation}

When $x_{1},x_{2},x_{3}$ $\longrightarrow \infty $, we obtain
\begin{eqnarray}
F(x_{1},x_{2},x_{3})
&=&\int_{x_{1}x_{2}x_{3}-z}^{(x_{1}+1)(x_{2}+1)(x_{3}+1)-z}\frac{1}{2\sqrt{u}%
}du  \nonumber \\[3mm]
&\geq &\frac{(x_{1}+1)(x_{2}+1)(x_{3}+1)-x_{1}x_{2}x_{3}}{2\sqrt{%
(x_{1}+1)(x_{2}+1)(x_{3}+1)-z}}  \nonumber \\[3mm]
&\geq &\frac{(x_{1}+1)(x_{2}+1)(x_{3}+1)-x_{1}x_{2}x_{3}}{2\sqrt{%
(x_{1}+1)(x_{2}+1)(x_{3}+1)}}  \nonumber \\[3mm]
&=&\frac{1}{2}\left[ \sqrt{\frac{x_{1}+1}{(x_{2}+1)(x_{3}+1)}}+\sqrt{\frac{%
(x_{2}+1)(x_{3}+1)}{x_{1}+1}}\right.\nonumber\\
&&\left.+\frac{x_{1}x_{2}+x_{1}x_{3}-1}{\sqrt{
(x_{1}+1)(x_{2}+1)(x_{3}+1)}}\right]  \nonumber \\[3mm]
&\geq
&1+\frac{x_{1}x_{2}+x_{1}x_{3}-1}{2\sqrt{(x_{1}+1)(x_{2}+1)(x_{3}+1)}}.
\end{eqnarray}
When one of the $x_{1}$, $x_{2}$ and $x_{3}$ goes to infinity, while the
rest keep finite, the proof can be done similarly.
Therefore, $F_{min}\geq 1$ holds for any fully
separable pure states. Finally, we have, from Lemma 1,
\begin{equation}
\label{ineq31} (\Delta K(\phi ))^{2}\geq 1
\end{equation}
for fully separable mixed states.

Actually inequality (\ref{ineq31}) is also true for any
bi-separable pure states. For example, for a $A|BC$ separable pure
state, we have
\begin{equation}
(\Delta K(\phi ))^{2}\geq \lbrack (\langle N_{a}+1\rangle \langle
(N_{b}+1)(N_{c}+1)\rangle -|\langle abc\rangle |^{2})^{1/2}
-(\langle N_{a}\rangle \langle N_{b}N_{c}\rangle -|\langle
abc\rangle |^{2})^{1/2}]^{2}.
\end{equation}%
Setting $x_{1}=\langle N_{a}\rangle$, $x_{2}=\langle N_{b}N_{c}\rangle$,
$x_{3}=\langle N_{b}\rangle $, $x_{4}=\langle N_{c}\rangle $ and $z=|\langle
abc\rangle |^{2},$ we define a function%
\begin{eqnarray}
G(x_{1},x_{2},x_{3},x_{4},z) &=&\sqrt{(x_{1}+1)(x_{2}+x_{3}+x_{4}+1)-z}-%
\sqrt{x_{1}x_{2}-z}  \nonumber \\
&\geq &\sqrt{(x_{1}+1)(x_{2}+1)-z}-\sqrt{x_{1}x_{2}-z},
\end{eqnarray}%
which is greater than or equal to 1 in the region of interest. From
lemma 1, we know that (\ref{ineq31}) is also valid for any
bi-partite separable mixed states. Therefore
violation of inequality (\ref{ineq31}) implies full entanglement.
This ends the proof.\qed

As an example we consider the GHZ state, $|GHZ\rangle =( |000\rangle
+ |111\rangle)/\sqrt{2}$. For this state, we have $(\Delta
K)^2=(\langle K^2\rangle)-(\langle K\rangle)^2=0,$ where
$K=K(\phi=0)$. Thus the GHZ state is fully entangled.

In analogy to (\ref{ineq10}), it is possible to find other
relations satisfied by separable states. For example, in the case of
pure fully separable
states we have that%
\begin{equation}
\label{ineq34} |\langle abc\rangle |=|\langle a\rangle \langle
b\rangle \langle c\rangle
|\leq \left[ \langle N_{a}\rangle \langle N_{b}\rangle \langle N_{c}\rangle %
\right] ^{1/2}.
\end{equation}%
However, (\ref{ineq34}) is not always true for some fully separable
mixed states. In fact, for a tripartite pure state, we can detect a
fully entangled state with respect to all three bipartite
decompositions. If all three bipartite decompositions are entangled,
the tripartite state is fully entangled.

Here is another criterion for detecting pure fully entangled states.
\begin{theorem}
If a pure state violates inequality (\ref{ineq34}) and satisfies the
following relations simultaneously,
\begin{eqnarray}
|\langle a\rangle \langle bc\rangle |^{2} &\leq&\langle N_{a}\rangle
\langle
N_{b}\rangle \langle N_{c}\rangle ,\label{111}\\
|\langle ab\rangle \langle c\rangle |^{2} &\leq&\langle N_{a}\rangle
\langle
N_{b}\rangle \langle N_{c}\rangle ,\\
|\langle ac\rangle \langle b\rangle |^{2} &\leq&\langle N_{a}\rangle
\langle N_{c}\rangle \langle N_{b}\rangle,
\end{eqnarray}
then the pure state is fully entangled.
\end{theorem}

 {\em Proof:} Without loss of
 generality, we suppose that a $A|BC$ separable state satisfies
 the above three inequalities and the inequality (\ref{ineq34}),
 $|\langle abc\rangle|>
 \left[\langle N_{a}\rangle \langle N_{b}\rangle \langle N_{c}\rangle
\right]^{1/2}$ simultaneously. However for a $A|BC$ separable state,
$|\langle abc\rangle|=|\langle a\rangle \langle bc\rangle |$.
This contradicts with (\ref{111}), $|\langle a\rangle \langle
bc\rangle |^{2} \leq \langle N_{a}\rangle \langle N_{b}\rangle
\langle N_{c}\rangle$. This ends the proof. \qed

We can extend these inequalities to general forms.
\begin{cor}
If a state satisfies the following relations simultaneously,
\begin{eqnarray}
&&\label{ineq39} |\langle a^{m}b^{n}c^{l}\rangle |> \left[ \langle
(a^{\dag })^{m}a^{m}\rangle \langle (b^{\dag })^{n}b^{n}\rangle
\langle (c^{\dag })^{l}c^{l}\rangle \right] ^{1/2}, \\
&&|\langle a^{m}b^{n}\rangle \langle c^{l}\rangle |^{2} \leq\langle
(a^{\dag })^{m}a^{m}\rangle \langle (b^{\dag })^{n}b^{n}\rangle
\langle (c^{\dag
})^{l}c^{l}\rangle ,\\
&&|\langle a^{m}\rangle \langle b^{n}c^{l}\rangle |^{2} \leq\langle
(a^{\dag })^{m}a^{m}\rangle \langle (b^{\dag })^{n}b^{n}\rangle
\langle (c^{\dag
})^{l}c^{l}\rangle ,\\
&&|\langle a^{m}c^{l}\rangle \langle b^{n}\rangle |^{2} \leq\langle
(a^{\dag })^{m}a^{m}\rangle \langle (c^{\dag })^{l}c^{l}\rangle
\langle (b^{\dag })^{n}b^{n}\rangle,
\end{eqnarray} where $m$, $n$ and $l$ are positive integers,
we can conclude that the state is fully entangled.
\end{cor}

Consider the state
\begin{equation}
|\psi \rangle =N_{-}(\alpha ,\beta ,\gamma )(|\alpha ,\beta ,\gamma \rangle
-|-\alpha ,-\beta ,-\gamma \rangle ),
\end{equation}%
where the normalization
\begin{equation}
N_{-}=\left[ 2\left( 1-e^{-2(|\alpha |^{2}+|\beta |^{2}+|\gamma
|^{2})}\right) \right] ^{-1/2},
\end{equation}
$|\alpha ,\beta ,\gamma \rangle$ is the coherent state. For this state
we have
\begin{eqnarray}
\label{ineq53} |\langle a^{\dag }(b)^{2}c\rangle |
&=&|\frac{1}{2\left( 1-e^{-2(|\alpha |^{2}+|\beta |^{2}+|\gamma
|^{2})}\right) }\left( \alpha ^{\ast }\langle \alpha ,\beta
,\gamma |+\alpha ^{\ast }\langle -\alpha ,-\beta ,-\gamma
|\right)   \nonumber \\
&&\left( \beta ^{2}\gamma |\alpha ,\beta ,\gamma \rangle +\beta ^{2}\gamma
|-\alpha ,-\beta ,-\gamma \rangle \right) |  \nonumber \\
&=&|\frac{2\alpha ^{\ast }\beta ^{2}\gamma (1+e^{-2(|\alpha
|^{2}+|\beta |^{2}+|\gamma |^{2})})}{2\left( 1-e^{-2(|\alpha
|^{2}+|\beta |^{2}+|\gamma |^{2})}\right) }| =|\alpha ^{\ast }\beta
^{2}\gamma |\coth (|\alpha |^{2}+|\beta |^{2}+|\gamma |^{2})
\end{eqnarray}%
and
\begin{eqnarray}
&&\label{ineq548} \left( \langle a^{\dag }ab^{\dag }b^{\dag
}bbc^{\dag }c\rangle \right) ^{1/2}\nonumber\\
&=&(\frac{1}{2\left( 1-e^{-2(|\alpha |^{2}+|\beta |^{2}+|\gamma
|^{2})}\right) }\left( \alpha ^{\ast }\beta ^{\ast }\beta ^{\ast
}\gamma ^{\ast }\langle \alpha ,\beta ,\gamma |-\alpha ^{\ast
}\beta ^{\ast }\beta ^{\ast }\gamma ^{\ast }\langle -\alpha
,-\beta ,-\gamma |\right) \nonumber
\\
&&\left( \alpha \beta ^{2}\gamma |\alpha ,\beta ,\gamma \rangle -\alpha
\beta ^{2}\gamma |-\alpha ,-\beta ,-\gamma \rangle \right) )^{1/2}  \nonumber
\\
&=&\left( \frac{2|\alpha |^{2}|\beta |^{4}|\gamma
|^{2}(1-e^{-2(|\alpha |^{2}+|\beta |^{2}+|\gamma |^{2})})}{2\left(
1-e^{-2(|\alpha |^{2}+|\beta |^{2}+|\gamma |^{2})}\right) }\right)
^{1/2} =|\alpha ||\beta |^{2}|\gamma |.
\end{eqnarray}
Eq.\ (\ref{ineq548}) is clearly less than Eq.\ (\ref{ineq53}) for all
nonzero parameters. Hence, the state is $A|BC$ entangled from inequality
(\ref{ineq17}). It is easily shown that for this state
$|a^{2}b^{\dag }c|^{2}>\langle a^{\dag }a^{\dag }aaN_{b}N_{c}\rangle
,$ $|a^{2}bc^{\dag }|^{2}>\langle N_{a}N_{b}c^{\dag }c^{\dag
}cc\rangle$. Therefore we conclude from  (\ref{ineq19}) and
(\ref{ineq10}) that this state is not only $AB|C$ entangled but
also $B|AC$ entangled: this state is fully entangled.

\section{multi-mode state}
The methods employed above for 3-mode states can be extended to
n-mode states. Consider $n$ modes whose annihilation operators are
$a_{1},a_{2},\cdots $ and $a_{n}$, respectively. For a state that is
separable between $m$-mode and $(n-m)$-mode, we have that
\begin{equation}
\label{ineq55} |\langle a_{1}a_{2}\cdots a_{m}a_{m+1}^{\dag
}\cdots a_{n}^{\dag }\rangle |^{2}\leq \langle N_{1}N_{2}\cdots
N_{n}\rangle ,
\end{equation}%
\begin{equation}
\label{ineq56} |\langle a_{1}a_{2}\cdots a_{m}a_{m+1}\cdots
a_{n}\rangle |^{2}\leq \langle N_{1}N_{2}\cdots N_{m}\rangle
\langle N_{m+1}\cdots N_{n}\rangle .
\end{equation}%
Therefore, if the inequalities (\ref{ineq55}) or
(\ref{ineq56}) are violated, the state is entangled between
$m$-mode and $(n-m)$-mode. We also have that for
a general $m|(n-m)$ separable state%
\begin{equation}
\label{ineq57} |\langle a_{1}^{l_{1}}a_{2}^{l_{2}}\cdots
a_{m}^{l_{m}}\cdots (a_{n}^{\dag })^{l_{n}}\rangle |^{2}\leq
\langle (a_{1}^{\dag })^{l_{1}}a_{1}^{l_{1}}(a_{2}^{\dag
})^{l_{2}}a_{2}^{l_{2}}\cdots (a_{m}^{\dag
})^{l_{m}}a_{m}^{l_{m}}\rangle ,
\end{equation}%
\begin{equation}
\label{ineq58} |\langle a_{1}^{l_{1}}a_{2}^{l_{2}}\cdots
a_{n}^{l_{n}}\rangle |^{2}\leq \langle (a_{1}^{\dag
})^{l_{1}}a_{1}^{l_{1}}(a_{2}^{\dag })^{l_{2}}a_{2}^{l_{2}}\cdots
(a_{m}^{\dag })^{l_{m}}a_{m}^{l_{m}}\rangle \langle (a_{m+1}^{\dag
})^{l_{m+1}}a_{m+1}^{l_{m+1}}\cdots (a_{m}^{\dag
})^{l_{m}}a_{m}^{l_{m}}\rangle .
\end{equation}%
Denote $K(\phi )=e^{i\phi }a_{1}a_{2}\cdots a_{n}+e^{-i\phi
}a_{1}^{\dag }a_{2}^{\dag }\cdots a_{n}^{\dag },$ we have that for a
fully separable state
\begin{equation}
\label{ineq54} (\Delta K(\phi ))^{2}\geq 1.
\end{equation}%
In analogy to formula (\ref{ineq31}), (\ref{ineq54}) holds for any
separable states. Hence if the inequality (\ref{ineq54}) is
violated, the multi-mode state must be fully entangled.

\section{conclusions}

We have studied inseparability conditions for multi-mode states by
presenting a series of inequalities. These conditions provide, in
principle, measurable tests of entanglement, in the sense that all
of the quantities appeared in the inequalities can be measured
experimentally. The results in section 2 (theorems 4-7) have simple forms and can easily applied.
They coincide with the ones in \cite{shchukin,peter,shchukin3} that can be
obtained by extending the methods used on non-positivity of
partial transposition (NPT) in \cite{shchukin,shchukin3}.
In many practical cases, such as teleportation network, we
firstly need to detect the fully entanglement of a state.
Our theorem 8-9 could be applied for such purpose, as the
quantities in the inequalities can be measured experimentally.
For dealing with the
entanglement of Gaussian states there are already very nice
results \cite{simon,lmd,serafini,ferraro}. Our theorems 8-9 give
some results which are derived for general states. Hence they
could be applied for detecting entanglement of non-Gaussian states.

The work is partly supported by
NKBRPC(2004CB318000) and NSFC projects (10375038, 90403018, 10675086)

\end{document}